\documentclass[preprint,12pt]{elsarticle}

\usepackage{amssymb}
\usepackage{amsmath}
\usepackage{hyperref}
\hypersetup{
     colorlinks = true,
     citecolor  = cyan,  
     linkcolor  = magenta 
}
\usepackage{color,soul}
\usepackage{subcaption}

\journal{Physics Letters A}

\begin{document}

\begin{frontmatter}



\title{Beyond Attraction: A Novel Approach to Repulsive Casimir-Lifshitz Forces using Heterogeneous Off-stoichiometry in Gapped Metals}


 \author[label1,label2]{S. Pal\corref{cor3}}
 \ead{palsubhojit429@gmail.com}
 \author[label3]{S. Osella}
 \author[label5,label2]{O. I. Malyi}
  \author[label2,label3]{M. Bostr\"om\corref{cor2}}
\ead{mathias.bostrom@ensemble3.eu}
 
\cortext[cor3,cor2]{Corresponding author} 
  \affiliation[label1]{organization={Universita degli Studi di Palermo,
Dipartimento di Fisica e Chimica--Emilio Segre}, addressline={Via Archirafi 36},
city={Palermo},postcode={90123},
             country={Italy}}

 \affiliation[label2]{organization={Centre of Excellence ENSEMBLE3},
             addressline={Wolczynska Str. 133},
             city={Warsaw},
             postcode={01-919},
             country={Poland}}

\affiliation[label3]{organization={Chemical and Biological Systems Simulation Lab, Centre of New Technologies, University of Warsaw},
addressline={Banacha 2C},
city={Warsaw},
postcode={02-097},
country={Poland}}

\affiliation[label5]{organization={Qingyuan Innovation Laboratory},city={Quanzhou},postcode={362801},country={China}}

\begin{abstract}
We uncover a physical system that enables a switch between attractive and repulsive Casimir forces when a Teflon surface interacts with a new form of quantum material (i.e., gapped metal) surface across different liquid media. {Past works usually suggest that entropic effects on the Casimir force (originating partly from its zero frequency contribution that has an approximately linear temperature dependence)} occur at high temperatures and/or large separations. We {propose a new way to achieve a zero-frequency Casimir effect, which reveals the potential for a Casimir force quantum switching within nanometer distances-a scale often thought to be unattainable.} Furthermore, the heterogeneous surface can have different phases (both metallic and insulating). Hence, our results introduce a new method to induce phase (stoichiometry)-controlled attraction-repulsion transitions and to achieve quantum levitation in a liquid medium by tuning the liquid environment. This study opens up new possibilities for manipulating the Casimir interactions between surfaces, forming the basis for advancements in nanoscale technology.
\end{abstract}


\begin{graphicalabstract}
  \protect\includegraphics[width=\columnwidth]{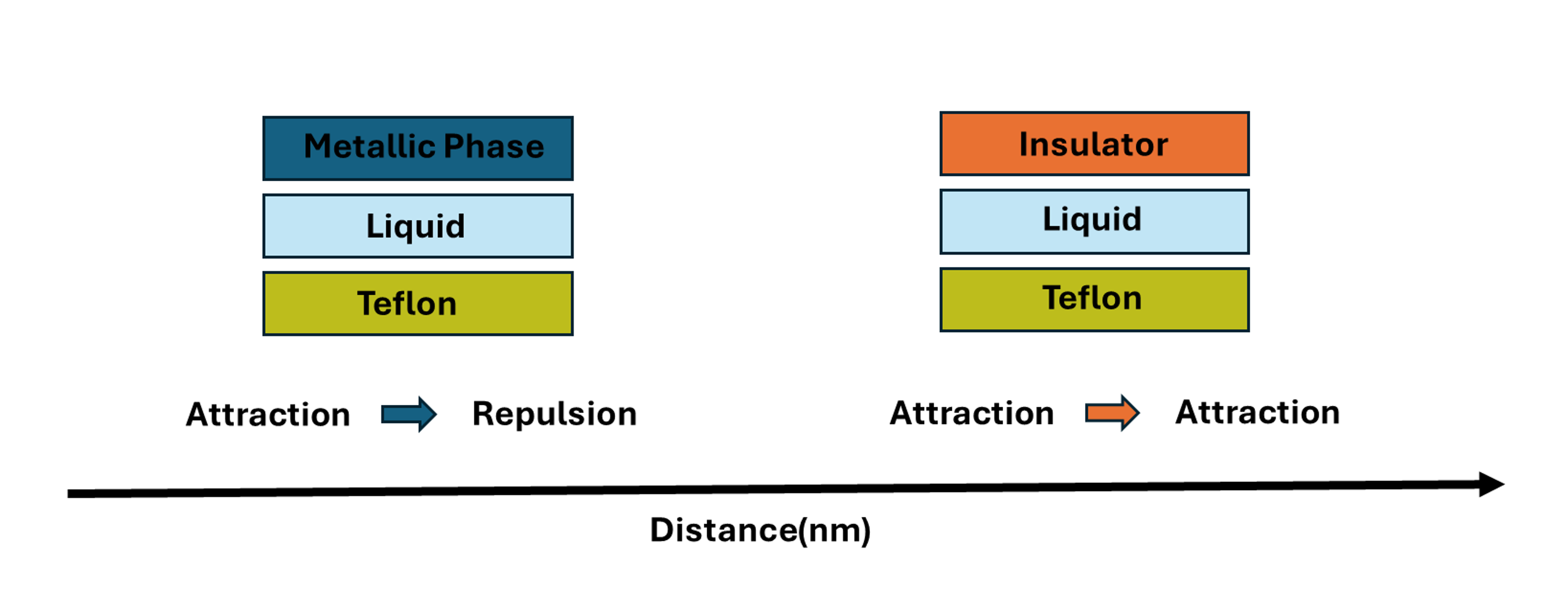}
\end{graphicalabstract}

\begin{highlights}
\item Casimir effect for heterogeneous phase change materials
\item Entropy effects on Casimir forces at nanometer scales
\item Proposed Casimir switch in fluids for nanotechnological applications 
\end{highlights}


\end{frontmatter}

\section{Introduction}
The potential to have repulsive Casimir-Lifshitz forces was implicit already in the original Lifshitz theory\,\cite{Dzya} and conclusively demonstrated in the 1970s\,\cite{AndSab,Haux,Rich71,Rich73}.
Past investigations demonstrating repulsive Casimir-Lifshitz forces have two important features: first, retardation effects are expected at relatively large distances (3-100 nm and beyond)\,\cite{PhysRevA.85.010701,esteso2015nanolevitation,VictoriaJAP2016,zhao2019stable,esteso2022effect}, and second, the zero frequency contribution to the Casimir force is {usually} only significant at much, much longer distances ($\gtrsim$ 100\,nm)\,\cite{Lamo1997,Bost2000,Bord,SushNP,klimchitskaya2019impact,VELICHKO2020100024}.
 {The engineered switching of Casimir force between attraction to repulsion has in the past been reported using topological insulators}\,\cite{PhysRevLett.106.020403,PhysRevB.88.085421},   {tin}\,\cite{PhysRevB.97.125421},  {and Vanadium dioxide}\,\cite{PhysRevB.101.104107,GE2022128392}. 
 {It is also worth mentioning that in a very recent paper}\,\cite{PhysRevApplied.21.044040} {, the Casimir force contribution from the
zero frequency term gave, similar to our results, an important contribution at a short separations smaller than 100 nm. This is because the the closeness of dielectric functions between solution and the coated nanofilms at high frequency, while the contrast of dielectric permittivities is high at zero frequency.}
 {As a positive  derivative of the free energy per unit area corresponds to repulsive pressure, the system in some, but not all, cases considered in the current work is unstable, i.e., the gap distance is a little larger than the zero-force position, and the top and bottom layers get far away, which is different from stable systems prepared by Zhao} {\it et al.}\,\cite{zhao2019stable}
 {Novel control of the switching from attractive to the repulsive forces by twist has in the past included exploiting layers of lithium iodate (LiIO$_3$), separated by a composite medium of nanoparticles dispersed in a liquid}\,\cite{twistinducedCasimirswitch2024}
A detailed review of the current state of Casimir effects and its versatile applications has been given by Klimchitskaya {\it et al.}\,\cite{RevModPhys.81.1827}.

In this work, we propose a novel approach based on recent developments in understanding the properties of gapped metals\cite{malyi2020false,khan2023optical}. These materials have a Fermi level located inside the conduction (or less often valence) bands, resulting in a high free carrier concentration in compounds with large internal band gaps between the principal band edges. Due to such a unique electronic structure, they could exhibit spontaneous off-stoichiometry. For instance, in the case of n-type gapped metals, the formation of acceptor cation vacancy can result in the decay of conducting electrons to the defect state, lowering defect formation energy\,\cite{malyi2019spontaneous}. Due to such unique physical phenomena, the properties of gapped metals are significantly distinct between different stoichiometries, often with a range of stable compounds that can be synthesized by tuning environmental conditions. Using the unique properties of these materials, we propose a novel approach that enables us to control the sign of the Casimir-Lifshitz free energy, by moving between surface regions with different optical properties to avoid stiction problems, equivalent to a simple method to turn a nanoscale switch on and off, based on off-stoichiometric properties of a gapped metal substrate interacting with a teflon nanosurface in fluids. This leads to observable sign changes, as well as retardation and entropic effects, at separations as small as a few nanometers.

\begin{figure}[!h]
    \centering
    \includegraphics[width=0.8\columnwidth]{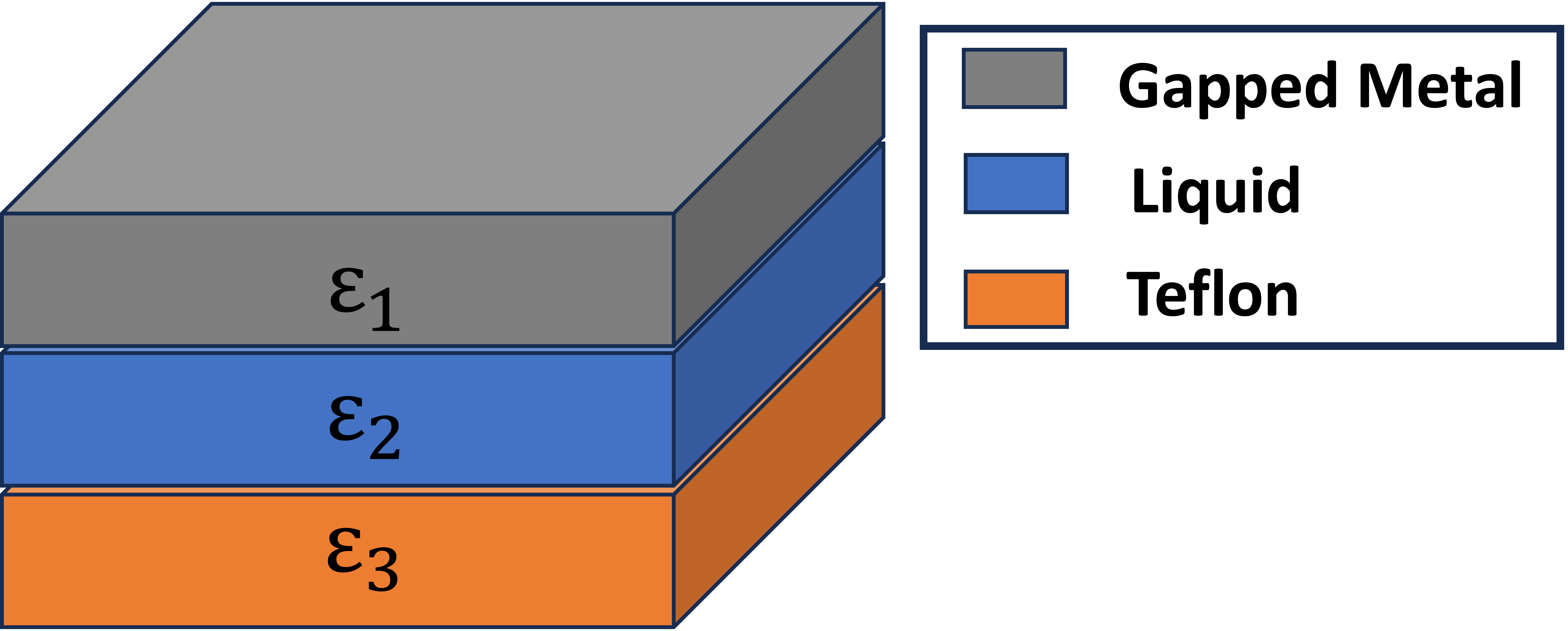}
    \caption{(Colors online) Schematic figure of the three-layer system involved here: gapped metal of infinite thickness and dielectric function $\varepsilon_1$, in contact with a liquid layer of dielectric function $\varepsilon_2$, of thickness $d$, which is in turn in contact with PTFE of dielectric function $\varepsilon_3$.}
    \label{fig:Scheme1}
\end{figure}

\begin{figure}[!h]
    \centering
    \includegraphics[width=0.8\columnwidth]{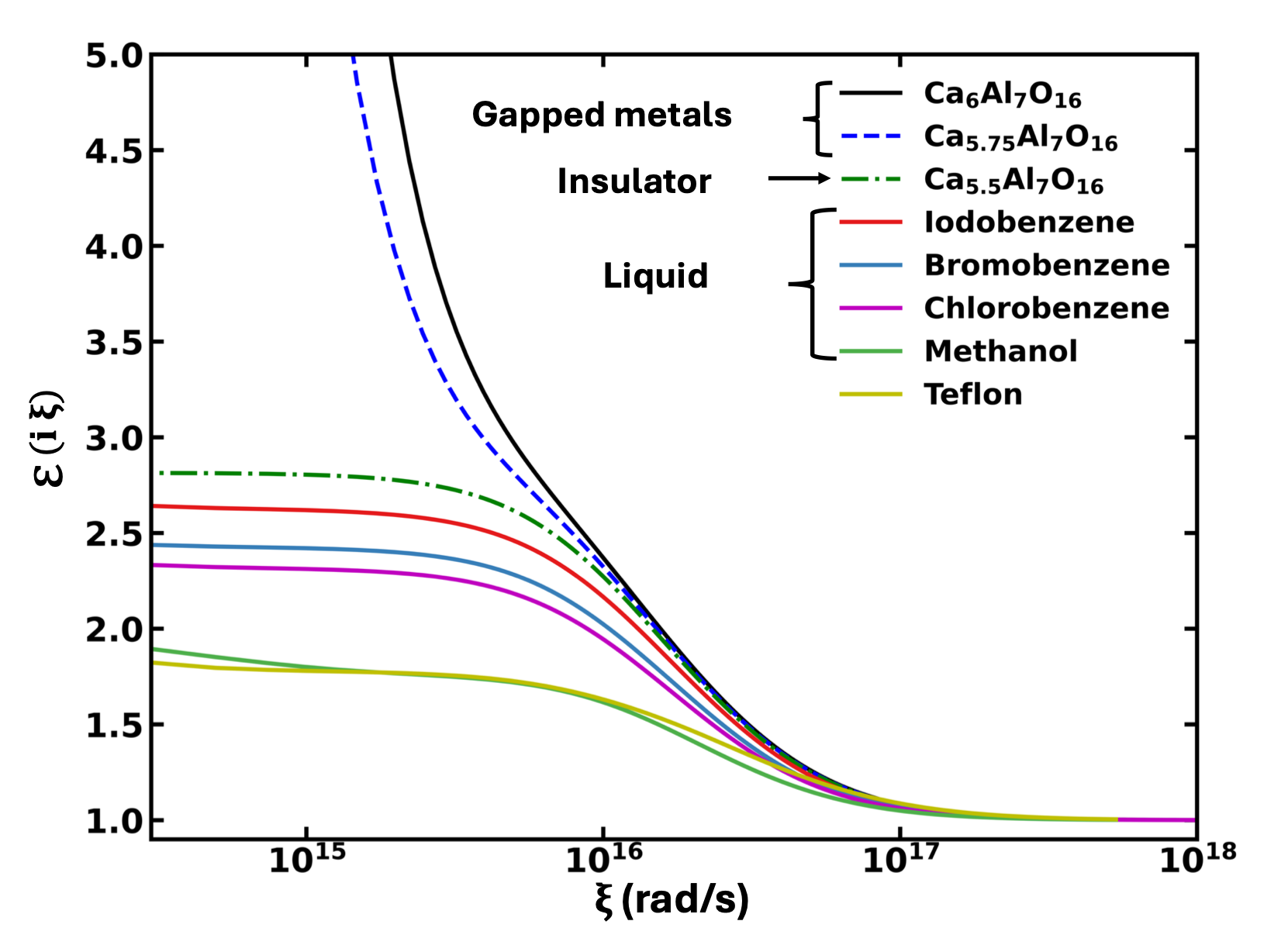}
    \caption{(Colors online) The dielectric functions at imaginary frequencies $\xi_{m}$ for different  Ca$_{6-x}$Al$_{7}$O$_{16}$ compounds (note that not all compounds are gapped metals as at high Ca vacancy concentration they become insulators),  liquids (Iodobenzene, Bromobenzene, Chlorobenzene and Methanol), and PTFE surface. Zero-frequency dielectric functions of all these materials and liquids are as follows, $\varepsilon_{\mathrm{Ca_{6}Al_{7}O_{16}}}(0)$ = 300.4216, $\varepsilon_{\mathrm{Ca_{5.75}Al_{7}O_{16}}}(0)$ = 58.4389, $\varepsilon_{\mathrm{Ca_{5.5}Al_{7}O_{16}}}(0)$ = 2.815, $\varepsilon_{\mathrm{Iodobenzene}}(0)$ = 4.6, 
    $\varepsilon_{\mathrm{Bromobenzene}}(0)$ = 5.37,
    $\varepsilon_{\mathrm{Chlorobenzene}}(0)$ = 5.75, $\varepsilon_{\mathrm{Methanol}}(0)$ = 32.9, $\varepsilon_{\mathrm{PTFE}}(0)$ = 2.1\,.}
    \label{fig:dielectricFig2}
\end{figure}

Notably, a method is shown where ``switching" can be achieved at a fixed distance by varying between different surface patches with different optical properties. We demonstrate that this can be achieved both at the extreme nanoscales where additional non-Casimir-Lifshitz forces originating from the fluid are expected to impact the measured force when the surfaces are a few nanometers apart and at the submicron scales. The retarded Casimir-Lifshitz interaction depends on an intricate balance between repulsive contributions from low frequencies ($\lesssim 10^{15}$ rad/s) and attractive contributions from higher frequency ranges, leading to the surprising importance of zero-frequency contributions at small distances, contrary to what is reported in the literature. Moreover, correlations of dielectric functions between different phases of gapped metals in the fully retarded theory influence a potential attractive-repulsive transition at 2 and 5.9\,nm for metallic phases of gapped metals, respectively. Past work described the repulsion between gapped metal surfaces and air bubbles\,\cite{BostromRizwanHarshanBrevikLiPerssonMalyi2023spontaneous} and attraction between a heterogeneous gapped metal surface and a gold surface\,\cite{PhysRevB.110.045424}. Herein, we explore a novel method for tuning attraction/repulsion transitions via combined retardation and entropic effects in previously unattainable nanolimits. One of the key elements to achieve repulsive Casimir interaction is the choice of intervening medium (i.e. liquid) between two interacting materials. In the case of repulsive Casimir interaction, the necessary conditions $\varepsilon_{1} > \varepsilon_{2} > \varepsilon_{3}$, where  $\varepsilon_{1}$, $\varepsilon_{3}$, and $\varepsilon_{2}$ represents the dielectric response of the two materials and the intervening liquid, respectively, must be satisfied. As we will demonstrate, exploiting the properties of gapped metals makes it possible to have both separation and phase change-induced attraction/repulsion transitions. Furthermore, from a fundamental point of view, it is of interest that retardation, the closeness of dielectric functions between liquids and Teflon, and zero frequency contribution are connected, enabling us to predict how to tune off-stoichiometrically controlled attraction/repulsion switches at unprecedented small separation ranges. To our knowledge, the striking dominance of the zero-frequency contribution at nanoscale separations and low temperatures has not been previously reported. Notably, this can be manipulated by varying the separation between metal and PTFE and interacting regions having different optical properties.


\section{Theory}

We consider the three-layer planar system reported in Fig.\,(\ref{fig:Scheme1}) as our model system,
where we calculate the Casimir-Lifshitz interaction free energy $F(d)$ between a planar gapped metal (medium 1), dielectric functions $\varepsilon_1(i \xi_m)$,  and a planar surface\,(medium 3) at a specific temperature $T=300$\,K from the calculated dielectric functions at discrete imaginary frequencies $\xi_{m}$ in Fig.\,(\ref{fig:dielectricFig2}). The medium 3 is considered to be made of Teflon ( {Polytetrafluoroethylene}, PTFE)\,\cite{Zwol2010}.  The intervening medium 2 is of the different liquids (Methanol, Bromobenzene, Iodobenzene, and Chlorobenzene) with dielectric functions $\varepsilon_2(i \xi_m)$. The dielectric functions of Teflon and the different liquids are taken from van Zwol and Palasantzas\,\cite{Zwol2010}.  We use DFT-calculated dielectric functions for the experimentally available Ca$_{6-x}$Al$_{7}$O$_{16}$ compounds\,\cite{BostromRizwanHarshanBrevikLiPerssonMalyi2023spontaneous,PhysRevB.110.045424}. We develop a parametrization of the average dielectric function for each of the considered compounds using an oscillator model: 

\begin{equation}
\varepsilon(i \xi)=1+\sum_j \frac{C_{j}}{1+ (\xi/\omega_j)^2}.
        \label{ParameteriseddielEq}
\end{equation}
Here, $\omega_j$ and $C_j$ represent the characteristic frequencies and the oscillator strength, respectively. The parameters can be obtained from Table\,(\ref{tab1}). Lebedew was the first to realize the connection between intermolecular forces and radiation processes in 1894\,\cite{Lebedev2}. The theory confirming this relation of optics with forces was presented in the classic paper by Lifshitz and co-workers\,\cite{Dzya}. The Casimir-Lifshitz interaction free energy  can be written in terms of the spectral function, $\mathrm{g}(\xi_{m})$, as\,\cite{NinhamParsegianWeiss1970,Ninhb,Ser2018},
 
\begin{equation}
\begin{aligned}
F(d) &= {\sum_{m=0}^\infty}{}^\prime \mathrm{g}(\xi_{m}) \\ &= 
    {\sum_{m=0}^\infty}{}^\prime \ \frac{k_BT}{2 \pi}  \int\limits_0^\infty dq\,q \sum_{\sigma}\ln(1- r_{\sigma}^{21}r_{\sigma}^{23}
 \mathrm e^{-2\kappa_2 d}) 
 \end{aligned}
  \label{LifFreeEnergy}
\end{equation}
with $F(d)=-A(d)/(12 \pi d^2)$ and where $\sigma=\rm TE,TM$, and the prime in the sum above indicates that the first term ($m$ = 0) has to be weighted by $1/2$. The Fresnel reflection coefficients between surfaces $i$ and $j$ for the transverse magnetic (TM) and transverse electric (TE) polarizations are given by
\begin{equation}
    r_{\rm TE}^{ij} = \frac{\kappa_i-\kappa_j}{\kappa_i+\kappa_j}\,;  \,\,\,\,\, r_{\rm TM}^{ij} = \frac{\varepsilon_j\kappa_i-\varepsilon_i \kappa_j}{\varepsilon_j \kappa_i+\varepsilon_i \kappa_j} \,. \label{eq:rtTETM}
\end{equation}
Here $\kappa_i= \sqrt{{q}^2+\varepsilon_i\xi_m^2/c^2}$, with $i=1,2,3$ and the Matsubara frequency being $\xi_m=2 \pi k T m/\hbar$, where $c$ and $k_B$ are the velocity of light in vacuum and the Boltzmann constant.
For our system, a balance between repulsive contributions in the low-frequency range with high-frequency attractive contributions leads to a possible sign change in the Casimir-Lifshitz interaction when comparing the non-retarded results with the fully retarded theory. The origin of this effect is due to the distance-dependent exponential of the spectral function $\mathrm{g}(\xi_{m})$ in Eq.\,(\ref{LifFreeEnergy}).
For fluid systems, additional short-range structural potentials can exist at interfaces. For liquids with isotropic or anisotropic molecules, the granularity near interfaces can give rise to attractive or repulsive oscillatory forces.
For up to at least 6 molecular layers, these forces can invalidate the use of the continuum Lifshitz theory\,\cite{QU200797,BONACCURSO2008107,luengo2022WaterIce}.
However, surface roughness can do two things: it can induce uncertainty in the distance and, potentially, smoothen out oscillations. Here it is important to recall that surface roughness can impact both the physics of nanoresonators\,\cite{Palasantzas2008_surfaceroughness} and the measured forces\,\cite{vanZwolRoughness2008}.  However, we will ignore structural and roughness contributions in our calculations. The focus is rather on exploring effects on the nanoscale that have previously been measured for many orders of magnitude larger separations\,\cite{zhao2019stable,LEE2002,Feiler2008,Munday2009}.

\section{Results}
 We consider here a three-layer Ca$_{6-x}$Al$_{7}$O$_{16}$--Methanol--PTFE interface. The metallic nature of gapped material is observed in Ca$_{6}$Al$_{7}$O$_{16}$ and Ca$_{5.75}$Al$_{7}$O$_{16}$, while Ca$_{5.5}$Al$_{7}$O$_{16}$ represents the insulating phase of gapped material. To explain our results, we first focus on the short-range limit. In the non-retarded limit ($c \to \infty$), the Fresnel reflection coefficients for transverse electric (TE) polarization approach zero, and the remaining contribution comes entirely from the transverse magnetic (TM) light polarization,
 \begin{equation}
    \mathrm{r}_{\rm NR}^{12} \mathrm{r}_{\rm NR}^{32}  = \frac{\varepsilon_1-\varepsilon_2  }{\varepsilon_1 +\varepsilon_2 } \times \frac{\varepsilon_3-\varepsilon_2 }{\varepsilon_3 +\varepsilon_2 }\,. \label{eq:rtTETM}
\end{equation}
Hence, the Hamaker constant (in the non-retarded limit) can be expressed as,
\begin{equation}
    A^{\text{NR}} = \frac{3}{2} k_{B}T {\sum_{m=0}^\infty}{}^\prime 
     \sum_{j=1}^{\infty} \Big[ r_{\text{NR}}^{21}(i \xi_{m} ) r_{\text{NR}}^{23}(i \xi_{m} )\Big]^j/j^3.
    \label{eqn:four}
\end{equation}

There are crossovers between the product of the reflection coefficients in the non-retarded limit for different stoichiometry of the gapped Ca$_{6-x}$Al$_{7}$O$_{16}$ metal surfaces in Fig.\,(\ref{fig:prod_reflection_coefficients}), that drives the possibility of attraction to repulsion transition in Casimir-Lifshitz interaction. 

\begin{figure}[!h]
    \centering
    \includegraphics[width=0.8\columnwidth]{Figure_3.png}
    \caption{(Colors online) Product of the reflection coefficients at imaginary frequencies $\xi_{m}$ in non-retarded limit for ranges of gapped metals,  Ca$_{6-x}$Al$_{7}$O$_{16}$.}
    \label{fig:prod_reflection_coefficients}
\end{figure}

Hence, in a broad frequency range $ \mathrm{r}_{\rm NR}^{12} \mathrm{r}_{\rm NR}^{32} > 0 $, the non-retarded Hamaker constant in Eq.\,(\ref{eqn:four}) is positive, corresponding to an attractive interaction. However, other frequency intervals produce repulsive contributions.
\begin{figure}[!htb]
    \centering
    \includegraphics[width=0.7\columnwidth]{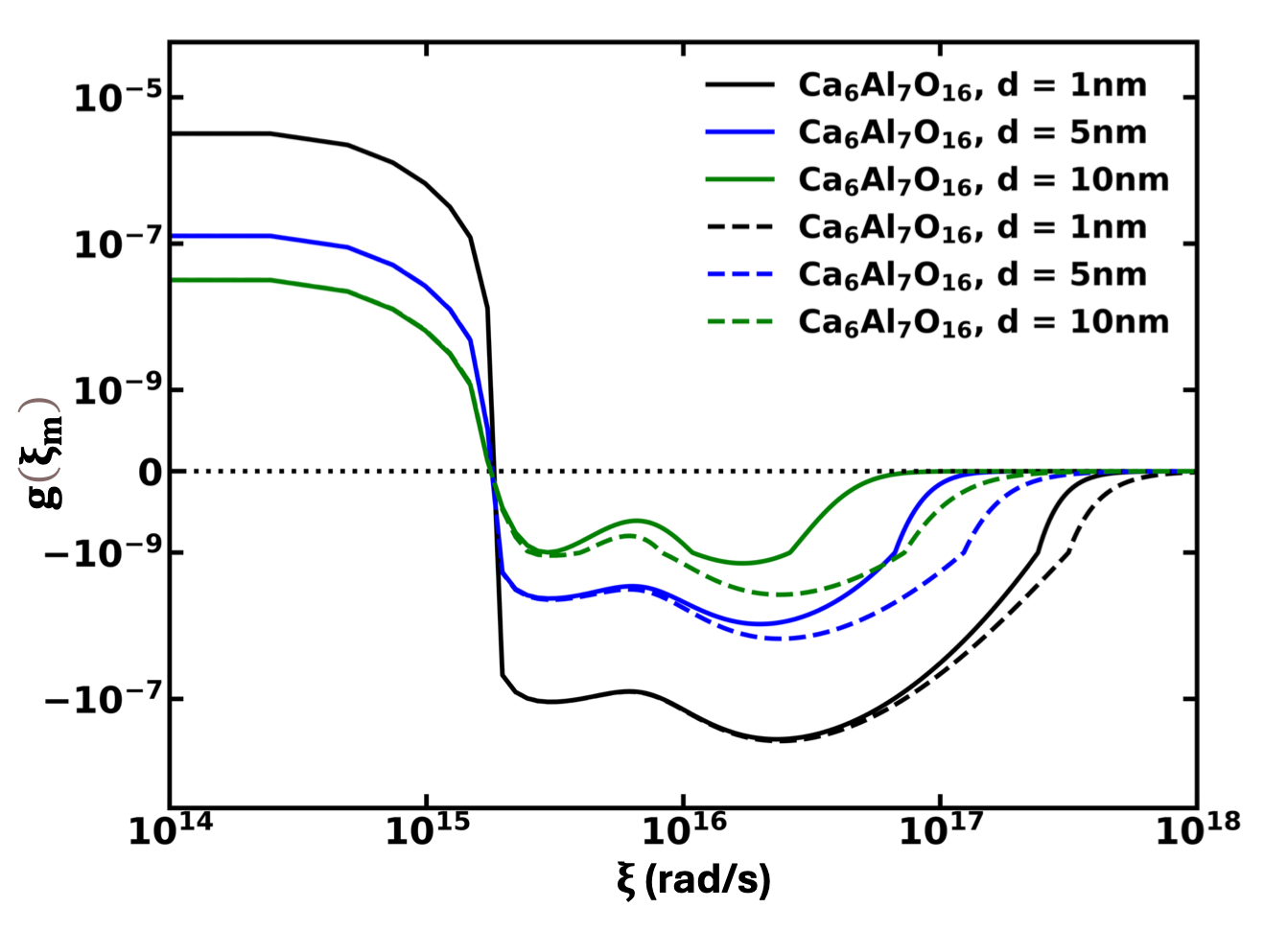}
    \caption{(Colors online) The spectral functions $\mathrm{g}(\xi_{m})$, defined in Eq.\,(\ref{LifFreeEnergy}), at different separations of liquid layer thickness in comparing the fully retarded and non-retarded limits in Ca$_{6-x}$Al$_{7}$O$_{16}$-Methanol-PTFE like systems. Solid lines indicate the retardation effect, while dashed lines represent the behavior of spectral functions in the non-retarded limit.}
    \label{fig:spectral_function}
\end{figure}

In Fig.\,(\ref{fig:spectral_function}), we present the decomposition of the Casimir-Lifshitz free energy into distance and frequency-dependent spectral functions,  $\mathrm{g}(\xi_{m})$.
At large separations of liquid layer thickness, retardation effect influences the contributions from all frequencies except for the extremely low ones. 
Ultimately, as it is well known\,\cite{Dzya,Bost2000,NinhamParsegianWeiss1970,PhysRevA.57.1870}, only the zero-frequency term survives. As the separation decreases, higher and higher frequencies play a role. 
As a result, the effect of retardation on repulsive contributions for the smallest separations is smaller since these originate from the low-frequency range. The key idea of this effect is a product of reflection coefficients (discussed in the previous part) along with a distance-dependent exponential factor $e^{-2 \sqrt{{q}^2 + \varepsilon_{2}\xi_m^2/c^2} d}$, 
meaning that significant contributions appear only when $\xi_{m} \sim \mathcal{O}(1/d)$ or smaller.

\begin{figure}
    \centering
    \includegraphics[width=0.7\linewidth]{Figure_5.png}
    \caption{(Colors online) Free energy vs. distance plot in Ca$_{6-x}$Al$_{7}$O$_{16}$ -- Methanol -- PTFE like system. Ca$_{6-x}$Al$_{7}$O$_{16}$ indicates different stoichiometry. Here red color dot and magenta color box denote maximum value of Casimir energy 
 for Ca$_{6}$Al$_{7}$O$_{16}$ and Ca$_{5.75}$Al$_{7}$O$_{16}$ are occurring at 2.9 nm  and 9.49 nm respectively. Here solid lines represent interaction energy in full retarded limit where dashed lines point the interaction energy in non-retarded limit.}
    \label{fig:freeEnergy_methanol}
\end{figure}

\begin{figure}[!htb]
    \centering
    \includegraphics[width=0.7\columnwidth]{Figure_6.png}
    \caption{ {(Colors online) The retarded Casimir-Lifshitz pressure vs. distance plot in Ca$_{6-x}$Al$_{7}$O$_{16}$ -- Methanol -- PTFE like system. Ca$_{6-x}$Al$_{7}$O$_{16}$ indicates different stoichiometry. Negative pressure corresponds to attraction.}   }
    \label{fig:pressure}
\end{figure}

In Fig.\,(\ref{fig:freeEnergy_methanol}), we notice that the non-retarded Casimir-Lifshitz interaction for different stoichiometry of  Ca$_{6-x}$Al$_{7}$O$_{16}$ is attractive in all separations, but there are reversals of signs for the retarded interaction around 2 nm and 5.9 nm for the metallic phases of the gapped metals Ca$_{6}$Al$_{7}$O$_{16}$ and Ca$_{5.75}$Al$_{7}$O$_{16}$, respectively. 
 {The sign reversal for pressure is shown in }Fig.\,\ref{fig:pressure}  {where a positive pressure corresponds to repulsion.}
This sign reversal can also be viewed from zero-frequency Hamaker constant in Table\,(\ref{HamakaerSubhojit}).
In order to demonstrate the long-range characteristics of various phases of gapped metals, the energy variation in the limit of long distances is illustrated in Fig.\,(\ref{fig:long-range_methanol}). It is observed that the metallic phases of gapped metals exhibit a repulsive effect even at long distances while the insulating phase displays an attractive interaction.

\begin{table*}[h]
\centering
\begin{tabular}{ p{2.8cm} | p{2.8cm} | p{2.8cm} | p{2.8cm} }
 ine
 ine
{modes ($\omega_j$)}  & \multicolumn{3}{c}{ Coefficients ($C_j$) for different Ca$_{6-x}$Al$_{7}$O$_{16}$ compounds}  \\
 ine
 & Ca$_{6}$Al$_{7}$O$_{16}$ & Ca$_{5.75}$Al$_{7}$O$_{16}$ & Ca$_{5.5}$Al$_{7}$O$_{16}$ \\
 ine
0.0206 & 58.9601 & 0.6494 & 0.0001 \\
 ine
0.0347 & 91.1774 & 1.797 & 0.0003 \\
 ine
0.0587 & 57.4068 & 5.2997 & 0.001 \\
 ine
0.1013 & 16.4729 & 15.4951 & 0.0221 \\
 ine
0.1996 & 73.0451 & 29.8463 & 0.3283 \\
 ine
0.3938 & 0.3949 & 2.1519 & 0.7511 \\
 ine
0.9556 & 0.0706 & 0.2519 & 0.4345 \\
 ine
2.2773 & 0.0987 & 0.0392 & 0.0 \\
 ine
6.4732 & 0.4594 & 0.2384 & 0.2279 \\
 ine
10.2048 & 0.7938 & 0.8189 & 0.0 \\
 ine
18.2421 & 0.3705 & 0.474 & 0.0486 \\
 ine
30.9018 & 0.1655 & 0.2388 & 0.0 \\
 ine
54.455 & 0.0059 & 0.0283 & 0.001 \\
 ine
 ine
\end{tabular}
\caption{Parametrization of the average dielectric function of continuous media, $\varepsilon(i\xi)$, at imaginary frequencies for Ca$_{6-x}$Al$_{7}$O$_{16}$ as calculated with first-principles calculations and a damping coefficient ($\Gamma$) set to 0.2 \,eV. The $\omega_j$ modes are given in $\rm eV$  {(1 eV=$1.5193 \times 10^{15}$\,rad/s)}. The largest difference between fitted and calculated $\varepsilon(i\xi)$ is 0.08\%. The density functional theory used to calculate the dielectric functions are described in our past works\,\cite{BostromRizwanHarshanBrevikLiPerssonMalyi2023spontaneous,PhysRevB.110.045424}.}
\label{tab1}
\end{table*}

\begin{table}[!h]
\centering
\begin{tabular}{|c|c|c|}
   ine
   Material & $A^{NR}$ & $A_{m=0}$ \\
     ine
      Ca$_{6}$Al$_{7}$O$_{16}$ & 0.004 eV &-0.012 eV\\
     ine
      Ca$_{5.75}$Al$_{7}$O$_{16}$ & 0.011 eV & -0.004 eV\\
     ine
     Ca$_{5.5}$Al$_{7}$O$_{16}$ & 0.033 eV &  0.016 eV\\
      ine
\end{tabular}
\caption{ The non-retarded Hamaker constant for gapped metals--methanol--PTFE combinations.}
\label{HamakaerSubhojit}
\end{table}
\begin{figure}[!htb]
    \centering
    \includegraphics[width=0.7\columnwidth]{Figure_7.png}
    \caption{(Colors online) Free energy vs.distance plot in Ca$_{6-x}$Al$_{7}$O$_{16}$  -- Methanol -- Teflon like system in the long-distance limit. Ca$_{6-x}$Al$_{7}$O$_{16}$ indicates different stoichiometry}.
    \label{fig:long-range_methanol}
\end{figure}

The dielectric functions are larger for the metallic phases than for the insulating phase in the low-frequency range leading to the enhancement of repulsive contributions. 
In this figure, there is a smooth transition between the attractive van der Waals regime and the long-range Casimir region around 2 nm and 5.9 nm for the metallic phases of the gapped metals. 

Remarkably, in most metallic cases the zero-frequency term dominates at 2.9 nm due to retardation and a strong cancellation of attractive and repulsive contributions. To the best of our knowledge, zero-frequency effects have never been predicted before, or observed, at such small distances at room temperature. 
 {For extremely short separations (less than $\sim$5-10\,nm), the use of wave vector independent permittivities can still be a reasonable first order approximation. This was shown by Bostr\"om and Sernelius using the random phase approximation as a model for a q-dependent permittivity for thin and thick metallic films.}\,\cite{PhysRevB.61.2204}. 
  {Nonlocality is of course expected to influence the results in the limit of very short separations.}\,\cite{Pendry2014,RevModPhys.88.045003}. 

\begin{figure}
    \centering
    \includegraphics[width=0.7\linewidth]{Figure_8.png}
    \caption{(Colors online) Decomposition of the contributions to the total non-retarded (non-ret) and retarded (ret) Casimir-Lifshitz free energy for Ca$_{6}$Al$_{7}$O$_{16}$--Methanol--PTFE like system.}
    \label{fig:energydecomposed}
\end{figure}
 
\begin{figure}[!htb]
    \centering 
    \includegraphics[width=0.8\columnwidth]{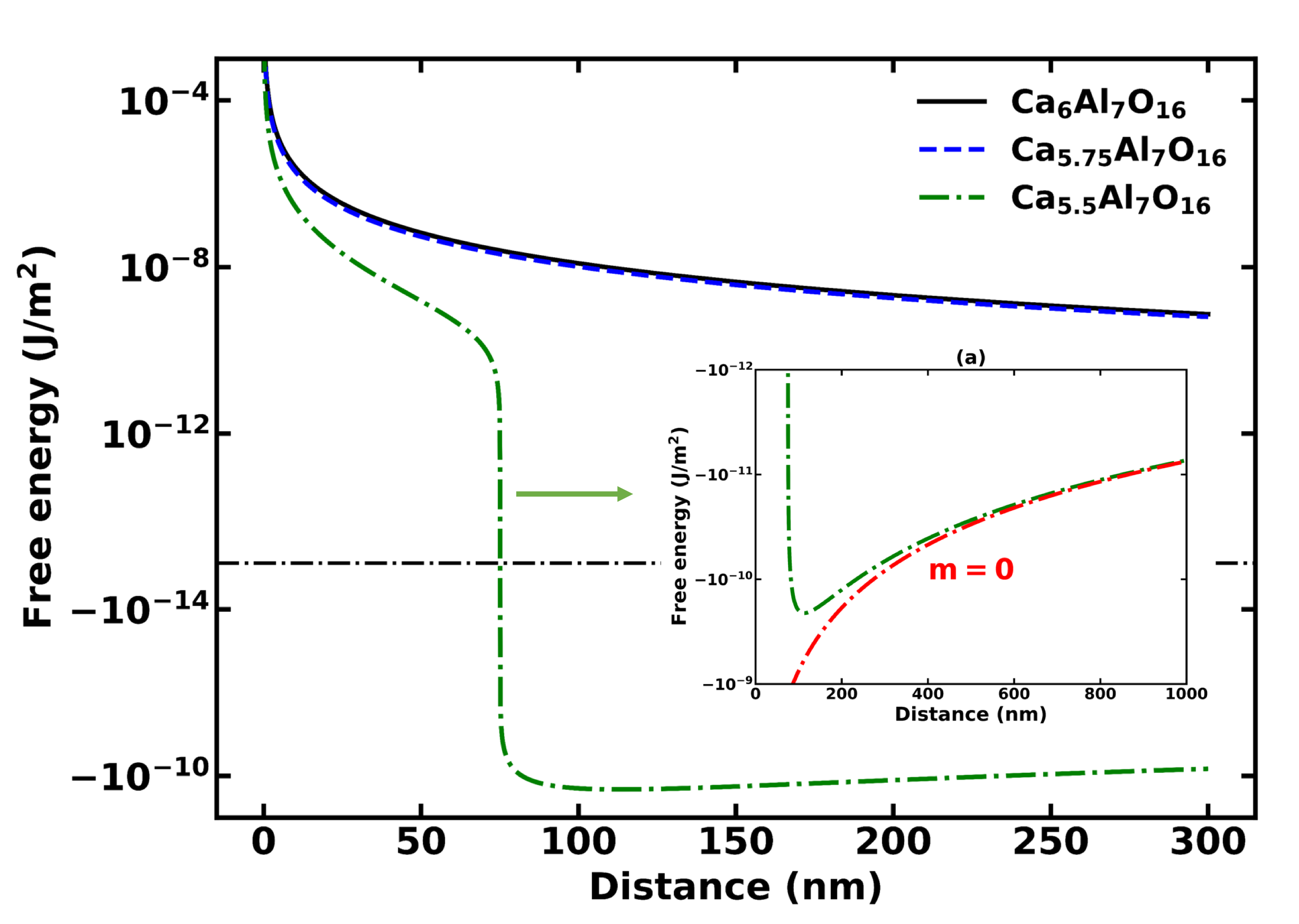}
    \caption{(Colors online) Free energy vs. distance plot in Ca$_{6-x}$Al$_{7}$O$_{16}$ --Iodobenzene-- PTFE like system. (a) In the inset, we illustrate the long-range characteristics of the Casimir-Lifshitz free energy, with an emphasis on the predominant contribution from zero-frequency term.  {The dotted black line represents zero free energy}.}
    \label{fig:freeEnergy_Iodobenzene}
\end{figure}

We observe that the long-range Casimir asymptote results in a repulsive effect originating from the presence of zero-frequency transverse magnetic (TM) modes and the influence of retardation. These results are consistent with Fig.\,(\ref{fig:spectral_function}). The contribution from TM modes is dominant for all distances but there is a small impact of TE modes, especially around 1.5\,nm\, where TM modes are canceling out in Fig.\,(\ref{fig:energydecomposed}). If we discard the impact of retardation, the contributions from these modes are absent, as shown in Fig.\,(\ref{fig:freeEnergy_methanol}).
We see exactly the opposite trends in the non-retarded limit in Fig.\,(\ref{fig:freeEnergy_Iodobenzene}) when the intervening medium is considered to be Iodobenzene instead of Methanol.  
 {The magnitudes of the dielectric functions for Iodobenzene, PTFE and insulating phase of gapped metal (Ca$_{5.5}$Al$_7$O$_{16}$) at finite and zero frequencies lead to a sign reversal around 75\,nm in Casimir energy as the distance varies. This gives rise to the predicted long-range attraction.}
The explanation is very straightforward: the repulsive interaction is linked to the short-range Hamaker constant crossing over to attraction when retardation effect mitigates the influence of finite frequency terms. The attraction is connected to the long-range zero frequency Hamaker constant. In contrast, for the other stoichiometries, the interaction remains repulsive across all distances. As shown in the inset of Fig.\,(\ref{fig:freeEnergy_Iodobenzene}), a mainly zero-frequency driven energy minima is predicted around\,0.11\,$\mu$m. This leads to the possibility of trapping a PTFE nanoparticle in iodobenzene near the surface of the gapped metal in the insulating phase. Similar analysis was also carried out using various liquids such as Bromobenzene and Chlorobenzene. Analogous patterns to those observed with Iodobenzene are identified, suggesting potential trapping of PTFE nanoparticles at the interface of the insulating layer of the gapped materials. 


\section{Conclusions}
 {Here, we have predicted that how moving across regions with different off-stoichiometry can be used to tailor the sign of the Casimir-Lifshitz force between the Teflon nanosurface and gapped metal surfaces separated by methanol. } Specifically, considering the range
of gapped metals Ca$_{6-x}$Al$_7$O$_{16}$, we demonstrate that controllable off-stoichiometry regions can be used to
tune the Lifshitz interaction from attraction to repulsion.  This behavior originates from the effect of off-stoichiometry on the dielectric function of gapped metals - the formation of cation vacancies
induces a reduction of free carriers and the Drude contribution to the dielectric function. We
show that the phase transition from insulator (Ca$_{5.5}$Al$_7$O$_{16}$) to metal (Ca$_{6}$Al$_7$O$_{16}$ and Ca$_{5.75}$Al$_7$O$_{16}$)
leads to a change in the force from attraction to repulsion at the nano to sub-micron scale driven
by retardation and caused by the difference in dielectric functions of the gapped metals.
The key point in our work is that a zero-frequency contribution and its interconnection with retardation and closeness of dielectric function of the liquids media and PTFE are the main origin for the sign change switching at unprecedented small distances. To the best of our knowledge, in the past purely zero-frequency effects have always been assumed to occur at much larger separations or at much higher temperatures\,\cite{PhysRevA.57.1870}. Here, we see its domination for separations as small as a few nanometers in the metallic case but not for the insulating phase.
While the dielectric functions used here do not have any temperature dependence (and hence can't be used to represent the real temperature-dependent picture), one can find that, e.g., 10\% changes in temperature, and the corresponding changes in the zero-frequency term, could lead to a measurable change in attraction-repulsion transition distances. With a change in background liquid media, it is also possible to have a zero-frequency Casimir-Lifshitz term leading to quantum levitation. In this way, using different liquids, we propose a new tool for nanoengineering via phase-transition-related quantum levitation in a liquid medium.

\section*{CRediT authorship contribution statement}
 
 {\bf{Subhojit Pal:}} Writing – review -\& editing, Writing – original draft, Software development, Investigation.
{\bf{Silvio Osella:}} Writing – review \& editing.
{\bf{Oleksandr Malyi:}} Writing – review \& editing, Materials modeling, Supervised computations, Computational resources acquisition. 
 {\bf{Mathias Bostr\"om:}} Writing – review -\& editing, Writing – original draft, Initiated and supervised the project,  Funding acquisition.

\section*{Declaration of Competing Interest}
The authors declare that they have no known competing financial interests or personal relationships that could have appeared to influence the work reported in this paper.

\section*{Acknowledgement}
SP, OIM, and MB's research contributions are part of the project No. 2022/47/P/ST3/01236 co-funded by the National Science Centre and the European Union's Horizon 2020 research and innovation programme under the Marie Sk{\l}odowska-Curie grant agreement No. 945339.
Institutional and infrastructural
support for the ENSEMBLE3 Centre of Excellence was
provided through the ENSEMBLE3 project (MAB/2020/14)
delivered within the Foundation for Polish Science International
Research Agenda Programme and cofinanced by the
European Regional Development Fund and the Horizon 2020
Teaming for Excellence initiative (Grant Agreement No.
857543), as well as the Ministry of Education and Science
initiative “Support for Centres of Excellence in Poland under
Horizon 2020” (MEiN/2023/DIR/3797). S.O. thanks the National Science Centre, Poland (grant no. UMO/2020/39/I/ST4/01446) and the ``Excellence Initiative - Research University" (IDUB) Program, Action I.3.3 - ``Establishment of the Institute for Advanced Studies (IAS)" for funding (grant no. UW/IDUB/2020/25).
We gratefully acknowledge Poland's high-performance computing infrastructure PLGrid (HPC Centers: ACK Cyfronet AGH) for providing computer facilities and support within computational grant no. PLG/2023/016228 and for awarding this project access to the LUMI supercomputer, owned by the EuroHPC Joint Undertaking, hosted by CSC (Finland) and the LUMI consortium through grant no. PLL/2023/4/016319.  M.B. gratefully acknowledges the kind hospitality at the University of Cagliari where this work was completed.







\begin{thebibliography}{10}
\expandafter\ifx\csname url\endcsname\relax
  \def\url#1{\texttt{#1}}\fi
\expandafter\ifx\csname urlprefix\endcsname\relax\def\urlprefix{URL }\fi
\expandafter\ifx\csname href\endcsname\relax
  \def\href#1#2{#2} \def\path#1{#1}\fi

\bibitem{Dzya}
I.~Dzyaloshinskii, E.~Lifshitz, L.~Pitaevskii, {The general theory of van der Waals forces}, Adv. Phys. 10~(38) (1961) 165--209.
\newblock \href {https://doi.org/10.1080/00018736100101281} {\path{doi:10.1080/00018736100101281}}.

\bibitem{AndSab}
C.~H. Anderson, E.~S. Sabisky, \href{https://link.aps.org/doi/10.1103/PhysRevLett.24.1049}{{Phonon Interference in Thin Films of Liquid Helium}}, Phys. Rev. Lett. 24 (1970) 1049--1052.
\newblock \href {https://doi.org/10.1103/PhysRevLett.24.1049} {\path{doi:10.1103/PhysRevLett.24.1049}}.
\newline\urlprefix\url{https://link.aps.org/doi/10.1103/PhysRevLett.24.1049}

\bibitem{Haux}
F.~Hauxwell, R.~H. Ottewill, \href{http://www.sciencedirect.com/science/article/pii/0021979770902080}{{A study of the surface of water by hydrocarbon adsorption}}, J. Colloid Int. Science 34~(4) (1970) 473 -- 479.
\newblock \href {https://doi.org/https://doi.org/10.1016/0021-9797(70)90208-0} {\path{doi:https://doi.org/10.1016/0021-9797(70)90208-0}}.
\newline\urlprefix\url{http://www.sciencedirect.com/science/article/pii/0021979770902080}

\bibitem{Rich71}
P.~Richmond, B.~W. Ninham, \href{http://www.sciencedirect.com/science/article/pii/0038109871904595}{{Calculations, using Lifshitz theory, of the height vs. thickness for vertical liquid helium films}}, Solid State Commun. 9~(13) (1971) 1045 -- 1047.
\newblock \href {https://doi.org/https://doi.org/10.1016/0038-1098(71)90459-5} {\path{doi:https://doi.org/10.1016/0038-1098(71)90459-5}}.
\newline\urlprefix\url{http://www.sciencedirect.com/science/article/pii/0038109871904595}

\bibitem{Rich73}
P.~Richmond, B.~W. Ninham, R.~H. Ottewill, \href{http://www.sciencedirect.com/science/article/pii/0021979773902439}{{A theoretical study of hydrocarbon adsorption on water surfaces using Lifshitz theory}}, J. Colloid Int. Sci. 45~(1) (1973) 69 -- 80.
\newblock \href {https://doi.org/https://doi.org/10.1016/0021-9797(73)90243-9} {\path{doi:https://doi.org/10.1016/0021-9797(73)90243-9}}.
\newline\urlprefix\url{http://www.sciencedirect.com/science/article/pii/0021979773902439}

\bibitem{PhysRevA.85.010701}
M.~Bostr\"om, B.~E. Sernelius, I.~Brevik, B.~W. Ninham, \href{https://link.aps.org/doi/10.1103/PhysRevA.85.010701}{Retardation turns the van der {Waals} attraction into a {Casimir} repulsion as close as 3 nm}, Phys. Rev. A 85 (2012) 010701.
\newblock \href {https://doi.org/10.1103/PhysRevA.85.010701} {\path{doi:10.1103/PhysRevA.85.010701}}.
\newline\urlprefix\url{https://link.aps.org/doi/10.1103/PhysRevA.85.010701}

\bibitem{esteso2015nanolevitation}
V.~Esteso, S.~Carretero-Palacios, H.~M{\'\i}guez, \href{https://pubs.acs.org/doi/epdf/10.1021/jp511851z}{Nanolevitation phenomena in real plane-parallel systems due to the balance between casimir and gravity forces}, J. Phys. Chem. C 119~(10) (2015) 5663--5670.
\newblock \href {https://doi.org/10.1021/jp511851z} {\path{doi:10.1021/jp511851z}}.
\newline\urlprefix\url{https://pubs.acs.org/doi/epdf/10.1021/jp511851z}

\bibitem{VictoriaJAP2016}
V.~Esteso, S.~Carretero-Palacios, H.~Miguez, \href{https://doi.org/10.1063/1.4945428}{{Effect of temperature variations on equilibrium distances in levitating parallel dielectric plates interacting through Casimir forces}}, Journal of Applied Physics 119~(14) (2016) 144301.
\newblock \href {http://arxiv.org/abs/https://doi.org/10.1063/1.4945428} {\path{arXiv:https://doi.org/10.1063/1.4945428}}, \href {https://doi.org/10.1063/1.4945428} {\path{doi:10.1063/1.4945428}}.
\newline\urlprefix\url{https://doi.org/10.1063/1.4945428}

\bibitem{zhao2019stable}
R.~Zhao, L.~Li, S.~Yang, W.~Bao, Y.~Xia, P.~Ashby, Y.~Wang, X.~Zhang, Stable {C}asimir equilibria and quantum trapping, Science 364~(6444) (2019) 984--987.
\newblock \href {https://doi.org/10.1126/science.aax0916} {\path{doi:10.1126/science.aax0916}}.

\bibitem{esteso2022effect}
V.~Esteso, S.~Carretero-Palacios, H.~M{\'\i}guez, Effect of spatial inhomogeneity on quantum trapping, J. Phys. Chem. Lett. 13~(20) (2022) 4513--4519.
\newblock \href {https://doi.org/10.1021/acs.jpclett.2c00807} {\path{doi:10.1021/acs.jpclett.2c00807}}.

\bibitem{Lamo1997}
S.~K. Lamoreaux, {Demonstration of the Casimir Force in the 0.6 to 6\,$\mu$\,m Range}, Phys. Rev. Lett. 78 (1997) 5.
\newblock \href {https://doi.org/10.1103/PhysRevLett.78.5} {\path{doi:10.1103/PhysRevLett.78.5}}.

\bibitem{Bost2000}
M.~Bostr\"om, B.~E. Sernelius, {Thermal Effects on the Casimir Force in the 0.1-5\,$\mu$\,m Range}, Phys. Rev. Lett. 84 (2000) 4757.
\newblock \href {https://doi.org/10.1103/PhysRevLett.84.4757} {\path{doi:10.1103/PhysRevLett.84.4757}}.

\bibitem{Bord}
M.~Bordag, B.~Geyer, G.~L. Klimchitskaya, V.~M. Mostepanenko, {Casimir Force at Both Nonzero Temperature and Finite Conductivity}, Phys. Rev. Lett. 85 (2000) 503.
\newblock \href {https://doi.org/10.1103/PhysRevLett.85.503} {\path{doi:10.1103/PhysRevLett.85.503}}.

\bibitem{SushNP}
A.~O. Sushkov, W.~J. Kim, D.~A.~R. Dalvit, S.~K. Lamoreaux, {Observation of the thermal Casimir force}, Nature Phys. 7 (2011) 230.
\newblock \href {https://doi.org/10.1038/nphys1909} {\path{doi:10.1038/nphys1909}}.

\bibitem{klimchitskaya2019impact}
G.~Klimchitskaya, V.~Mostepanenko, E.~Nepomnyashchaya, E.~Velichko, \href{https://link.aps.org/doi/10.1103/PhysRevB.99.045433}{Impact of magnetic nanoparticles on the {C}asimir pressure in three-layer systems}, Phys. Rev. B 99~(4) (2019) 045433.
\newblock \href {https://doi.org/10.1103/PhysRevB.99.045433} {\path{doi:10.1103/PhysRevB.99.045433}}.
\newline\urlprefix\url{https://link.aps.org/doi/10.1103/PhysRevB.99.045433}

\bibitem{VELICHKO2020100024}
E.~N. Velichko, G.~L. Klimchitskaya, E.~N. Nepomnyashchaya, \href{https://www.sciencedirect.com/science/article/pii/S1674862X20300215}{Casimir effect in optoelectronic devices using ferrofluids}, J. Electron. Sci. Technol/ 18~(1) (2020) 100024.
\newblock \href {https://doi.org/https://doi.org/10.1016/j.jnlest.2020.100024} {\path{doi:https://doi.org/10.1016/j.jnlest.2020.100024}}.
\newline\urlprefix\url{https://www.sciencedirect.com/science/article/pii/S1674862X20300215}

\bibitem{PhysRevLett.106.020403}
A.~G. Grushin, A.~Cortijo, \href{https://link.aps.org/doi/10.1103/PhysRevLett.106.020403}{Tunable casimir repulsion with three-dimensional topological insulators}, Phys. Rev. Lett. 106 (2011) 020403.
\newblock \href {https://doi.org/10.1103/PhysRevLett.106.020403} {\path{doi:10.1103/PhysRevLett.106.020403}}.
\newline\urlprefix\url{https://link.aps.org/doi/10.1103/PhysRevLett.106.020403}

\bibitem{PhysRevB.88.085421}
W.~Nie, R.~Zeng, Y.~Lan, S.~Zhu, \href{https://link.aps.org/doi/10.1103/PhysRevB.88.085421}{Casimir force between topological insulator slabs}, Phys. Rev. B 88 (2013) 085421.
\newblock \href {https://doi.org/10.1103/PhysRevB.88.085421} {\path{doi:10.1103/PhysRevB.88.085421}}.
\newline\urlprefix\url{https://link.aps.org/doi/10.1103/PhysRevB.88.085421}

\bibitem{PhysRevB.97.125421}
M.~Bostr\"om, M.~Dou, O.~I. Malyi, P.~Parashar, D.~F. Parsons, I.~Brevik, C.~Persson, \href{https://link.aps.org/doi/10.1103/PhysRevB.97.125421}{Fluid-sensitive nanoscale switching with quantum levitation controlled by $\ensuremath{\alpha}$-sn/$\ensuremath{\beta}$-sn phase transition}, Phys. Rev. B 97 (2018) 125421.
\newblock \href {https://doi.org/10.1103/PhysRevB.97.125421} {\path{doi:10.1103/PhysRevB.97.125421}}.
\newline\urlprefix\url{https://link.aps.org/doi/10.1103/PhysRevB.97.125421}

\bibitem{PhysRevB.101.104107}
L.~Ge, X.~Shi, Z.~Xu, K.~Gong, \href{https://link.aps.org/doi/10.1103/PhysRevB.101.104107}{Tunable casimir equilibria with phase change materials: From quantum trapping to its release}, Phys. Rev. B 101 (2020) 104107.
\newblock \href {https://doi.org/10.1103/PhysRevB.101.104107} {\path{doi:10.1103/PhysRevB.101.104107}}.
\newline\urlprefix\url{https://link.aps.org/doi/10.1103/PhysRevB.101.104107}

\bibitem{GE2022128392}
L.~Ge, X.~Shi, \href{https://www.sciencedirect.com/science/article/pii/S0375960122004741}{Thermal hysteresis of casimir suspensions enabled by vanadium dioxide}, Physics Letters A 450 (2022) 128392.
\newblock \href {https://doi.org/https://doi.org/10.1016/j.physleta.2022.128392} {\path{doi:https://doi.org/10.1016/j.physleta.2022.128392}}.
\newline\urlprefix\url{https://www.sciencedirect.com/science/article/pii/S0375960122004741}

\bibitem{PhysRevApplied.21.044040}
L.~Ge, K.~Liu, K.~Gong, R.~Podgornik, \href{https://link.aps.org/doi/10.1103/PhysRevApplied.21.044040}{Fabry-p\'erot nanocavities controlled by casimir forces in electrolyte solutions}, Phys. Rev. Appl. 21 (2024) 044040.
\newblock \href {https://doi.org/10.1103/PhysRevApplied.21.044040} {\path{doi:10.1103/PhysRevApplied.21.044040}}.
\newline\urlprefix\url{https://link.aps.org/doi/10.1103/PhysRevApplied.21.044040}

\bibitem{twistinducedCasimirswitch2024}
Y.~Hu, X.~Wu, H.~Liu, W.~Ge, J.~Zhang, X.~Huang, \href{https://doi.org/10.1021/acsphotonics.4c00129}{Twist-induced casimir attractive-repulsive transition based on lithium iodate}, ACS Photonics 11~(5) (2024) 1998--2006.
\newblock \href {https://doi.org/10.1021/acsphotonics.4c00129} {\path{doi:10.1021/acsphotonics.4c00129}}.
\newline\urlprefix\url{https://doi.org/10.1021/acsphotonics.4c00129}

\bibitem{RevModPhys.81.1827}
G.~L. Klimchitskaya, U.~Mohideen, V.~M. Mostepanenko, \href{https://link.aps.org/doi/10.1103/RevModPhys.81.1827}{{The Casimir force between real materials: Experiment and theory}}, Rev. Mod. Phys. 81 (2009) 1827--1885.
\newblock \href {https://doi.org/10.1103/RevModPhys.81.1827} {\path{doi:10.1103/RevModPhys.81.1827}}.
\newline\urlprefix\url{https://link.aps.org/doi/10.1103/RevModPhys.81.1827}

\bibitem{malyi2020false}
O.~I. Malyi, A.~Zunger, \href{https://pubs.aip.org/aip/apr/article/7/4/041310/831966/False-metals-real-insulators-and-degenerate-gapped}{False metals, real insulators, and degenerate gapped metals}, Appl. Phys. Rev. 7~(4) (2020) 041310.
\newblock \href {https://doi.org/10.1063/5.0015322} {\path{doi:10.1063/5.0015322}}.
\newline\urlprefix\url{https://pubs.aip.org/aip/apr/article/7/4/041310/831966/False-metals-real-insulators-and-degenerate-gapped}

\bibitem{khan2023optical}
M.~R. Khan, H.~R. Gopidi, O.~I. Malyi, \href{https://doi.org/10.1063/5.0153382}{Optical properties and electronic structures of intrinsic gapped metals: Inverse materials design principles for transparent conductors}, Appl. Phys. Lett. 123~(6) (2023).
\newline\urlprefix\url{https://doi.org/10.1063/5.0153382}

\bibitem{malyi2019spontaneous}
O.~I. Malyi, M.~T. Yeung, K.~R. Poeppelmeier, C.~Persson, A.~Zunger, \href{https://www.sciencedirect.com/science/article/pii/S2590238519300396}{Spontaneous non-stoichiometry and ordering in degenerate but gapped transparent conductors}, Matter 1~(1) (2019) 280--294.
\newline\urlprefix\url{https://www.sciencedirect.com/science/article/pii/S2590238519300396}

\bibitem{BostromRizwanHarshanBrevikLiPerssonMalyi2023spontaneous}
M.~Bostr\"om, M.~R. Khan, H.~R. Gopidi, I.~Brevik, Y.~Li, C.~Persson, O.~I. Malyi, {Tuning the Casimir-Lifshitz force with gapped metals}, Phys. Rev. B 108 (2023) 165306.
\newblock \href {https://doi.org/10.1103/PhysRevB.108.165306} {\path{doi:10.1103/PhysRevB.108.165306}}.

\bibitem{PhysRevB.110.045424}
M.~Bostr\"om, S.~Pal, H.~R. Gopidi, S.~Osella, A.~Gholamhosseinian, G.~Palasantzas, O.~I. Malyi, \href{https://link.aps.org/doi/10.1103/PhysRevB.110.045424}{{Casimir-Lifshitz} force variations across heterogeneous gapped metal surfaces}, Phys. Rev. B 110 (2024) 045424.
\newblock \href {https://doi.org/10.1103/PhysRevB.110.045424} {\path{doi:10.1103/PhysRevB.110.045424}}.
\newline\urlprefix\url{https://link.aps.org/doi/10.1103/PhysRevB.110.045424}

\bibitem{Zwol2010}
P.~J. van Zwol, G.~Palasantzas, \href{https://link.aps.org/doi/10.1103/PhysRevA.81.062502}{{Repulsive Casimir forces between solid materials with high-refractive-index intervening liquids}}, Phys. Rev. A 81 (2010) 062502.
\newblock \href {https://doi.org/10.1103/PhysRevA.81.062502} {\path{doi:10.1103/PhysRevA.81.062502}}.
\newline\urlprefix\url{https://link.aps.org/doi/10.1103/PhysRevA.81.062502}

\bibitem{Lebedev2}
P.~Lebedew, {Ueber die mechanische Wirkung der Wellen auf ruhende Resonatoren. I. Electromagnetische Wellen}, Ann. der Phys. 288~(8) (1894) 621--640.
\newblock \href {https://doi.org/10.1002/andp.18942880803} {\path{doi:10.1002/andp.18942880803}}.

\bibitem{NinhamParsegianWeiss1970}
B.~W. Ninham, V.~A. Parsegian, G.~H. Weiss, {On the macroscopic theory of temperature dependent van der Waals forces}, J. Stat. Phys. 2 (1970) 323.
\newblock \href {https://doi.org/doi.org/10.1007/BF01020441} {\path{doi:doi.org/10.1007/BF01020441}}.

\bibitem{Ninhb}
B.~W. Ninham, P.~Lo~Nostro, Molecular Forces and Self Assembly in Colloid, Nano Sciences and Biology, Cambridge University Press, Cambridge, 2010.
\newblock \href {https://doi.org/10.1017/CBO9780511811531} {\path{doi:10.1017/CBO9780511811531}}.

\bibitem{Ser2018}
B.~E. Sernelius, \href{https://www.springer.com/us/book/9783319998305}{{Fundamentals of van der Waals and Casimir Interactions}}, Springer Series on Atomic, Optical, and Plasma Physics, Springer International Publishing, Switzerland, 2018.
\newblock \href {https://doi.org/10.1007/978-3-319-99831-2} {\path{doi:10.1007/978-3-319-99831-2}}.
\newline\urlprefix\url{https://www.springer.com/us/book/9783319998305}

\bibitem{QU200797}
D.~Qu, G.~Brotons, V.~Bosio, A.~Fery, T.~Salditt, D.~Langevin, R.~{von Klitzing}, Interactions across liquid thin films, Colloids and Surfaces A: Physicochemical and Engineering Aspects 303~(1) (2007) 97--109.
\newblock \href {https://doi.org/https://doi.org/10.1016/j.colsurfa.2007.03.055} {\path{doi:https://doi.org/10.1016/j.colsurfa.2007.03.055}}.

\bibitem{BONACCURSO2008107}
E.~Bonaccurso, M.~Kappl, H.-J. Butt, Thin liquid films studied by atomic force microscopy, Current Opinion in Colloid \& Interface Science 13~(3) (2008) 107--119.
\newblock \href {https://doi.org/https://doi.org/10.1016/j.cocis.2007.11.010} {\path{doi:https://doi.org/10.1016/j.cocis.2007.11.010}}.

\bibitem{luengo2022WaterIce}
J.~Luengo-Marquez, F.~Izquierdo-Ruiz, L.~G. MacDowell, Intermolecular forces at ice and water interfaces: Premelting, surface freezing, and regelation, J. Chem. Phys. 157~(4) (2022) 044704.
\newblock \href {https://doi.org/10.1063/5.0097378} {\path{doi:10.1063/5.0097378}}.

\bibitem{Palasantzas2008_surfaceroughness}
G.~Palasantzas, \href{https://doi.org/10.1063/1.2874790}{{Surface roughness influence on the quality factor of high frequency nanoresonators}}, Journal of Applied Physics 103~(4) (2008) 046106.
\newblock \href {https://doi.org/10.1063/1.2874790} {\path{doi:10.1063/1.2874790}}.
\newline\urlprefix\url{https://doi.org/10.1063/1.2874790}

\bibitem{vanZwolRoughness2008}
P.~J. van Zwol, G.~Palasantzas, M.~van~de Schootbrugge, J.~T.~M. de~Hosson, V.~S.~J. Craig, \href{https://doi.org/10.1021/la800664f}{Roughness of microspheres for force measurements}, Langmuir 24~(14) (2008) 7528--7531.
\newblock \href {https://doi.org/10.1021/la800664f} {\path{doi:10.1021/la800664f}}.
\newline\urlprefix\url{https://doi.org/10.1021/la800664f}

\bibitem{LEE2002}
S.~Lee, W.~M. Sigmund, {AFM study of repulsive van der Waals forces between Teflon AF$^{TM}$ thin film and silica or alumina}, Colloid. Surf. A: Physicochem. Eng. Asp. 204~(1) (2002) 43 -- 50.
\newblock \href {https://doi.org/https://doi.org/10.1016/S0927-7757(01)01118-9} {\path{doi:https://doi.org/10.1016/S0927-7757(01)01118-9}}.

\bibitem{Feiler2008}
A.~A. Feiler, L.~Bergstr{\"o}m, M.~W. Rutland, Superlubricity using repulsive van der {W}aals forces, Langmuir 24 (2008) 2274 -- 2276.
\newblock \href {https://doi.org/https://doi.org/10.1021/la7036907} {\path{doi:https://doi.org/10.1021/la7036907}}.

\bibitem{Munday2009}
J.~N. Munday, F.~Capasso, V.~A. Parsegian, Measured long-range repulsive casimir--lifshitz forces, Nature 457~(7226) (2009) 170--173.

\bibitem{PhysRevA.57.1870}
B.~W. Ninham, J.~Daicic, {Lifshitz theory of Casimir forces at finite temperature}, Phys. Rev. A 57 (1998) 1870--1880.
\newblock \href {https://doi.org/10.1103/PhysRevA.57.1870} {\path{doi:10.1103/PhysRevA.57.1870}}.

\bibitem{PhysRevB.61.2204}
M.~Bostr\"om, B.~E. Sernelius, \href{https://link.aps.org/doi/10.1103/PhysRevB.61.2204}{Fractional van der waals interaction between thin metallic films}, Phys. Rev. B 61 (2000) 2204--2210.
\newblock \href {https://doi.org/10.1103/PhysRevB.61.2204} {\path{doi:10.1103/PhysRevB.61.2204}}.
\newline\urlprefix\url{https://link.aps.org/doi/10.1103/PhysRevB.61.2204}

\bibitem{Pendry2014}
Y.~Luo, R.~Zhao, J.~Pendry, \href{https://doi.org/10.1073/pnas.1420551111}{van der waals interactions at the nanoscale: The effects of nonlocality}, Proc. Natl. Acad. Sci. U.S.A. 111~(52) (2014) 18422--18427.
\newblock \href {https://doi.org/https://doi.org/10.1073/pnas.1420551111} {\path{doi:https://doi.org/10.1073/pnas.1420551111}}.
\newline\urlprefix\url{https://doi.org/10.1073/pnas.1420551111}

\bibitem{RevModPhys.88.045003}
L.~M. Woods, D.~A.~R. Dalvit, A.~Tkatchenko, P.~Rodriguez-Lopez, A.~W. Rodriguez, R.~Podgornik, \href{https://link.aps.org/doi/10.1103/RevModPhys.88.045003}{Materials perspective on casimir and van der waals interactions}, Rev. Mod. Phys. 88 (2016) 045003.
\newblock \href {https://doi.org/10.1103/RevModPhys.88.045003} {\path{doi:10.1103/RevModPhys.88.045003}}.
\newline\urlprefix\url{https://link.aps.org/doi/10.1103/RevModPhys.88.045003}

\end{thebibliography}






\end{document}